\documentclass[twocolumn,amsmath,amssymb,floatfix,pra,showpacs,eqsecnum,footinbib,superscriptaddress]{revtex4}
\usepackage{amsmath,amssymb,natbib,bm,graphicx,url,psfrag, tikz}

\newtheorem{thm}{Theorem}

\def\NP{{\text{NP}}}
\def\tr{{\text{tr}}}
\def\T{{\text{T}}}

\begin{document}

\title{Entanglement-assisted zero-error capacity is upper bounded by the Lov\'{a}sz theta function}

\author{Salman Beigi}
\affiliation{Institute for Quantum Information, California Institute of
Technology, Pasadena, California, USA}

%\date{September 26, 2007}

\begin{abstract}
The zero-error capacity of a classical channel is expressed in terms of the independence number of some graph and its tensor powers. This quantity is hard to compute even for small graphs such as the cycle of length seven, so upper bounds such as the Lov\'{a}sz theta function play an important role in zero-error communication. In this paper, we show that the Lov\'{a}sz theta function is an upper bound on the zero-error capacity even in the presence of entanglement between the sender and receiver.
\end{abstract}

\pacs{03.67.Ac, 03.67.Bg, 89.70.Kn}

\maketitle

\section{Introduction}\label{sec:intro}

To provide an arbitrary small amount of error for transmitting information via a communication channel, the number of channel uses or equivalently the length of codewords should tend to infinity, which results in inefficient encoding and decoding processes. To overcome this problem Shannon \cite{shannon} defined the zero-error channel capacity as the maximum rate of information that can be sent through the channel with no error. Later it was shown by Shannon, Gallager, and Berlekamp \cite{sgb} that even the rate at which the probability of error in the usual definition of channel capacity decreases is related to the zero-error capacity, turning this quantity into an important notion in information theory \cite{korner}.

To compute the zero-error capacity, the exact distribution of the output of the channel under a certain input is not important; what matters is whether the probability of receiving an output is zero or not. Indeed, to encode $m$ messages into codewords of \emph{length one} and transmitting them with zero-error we should find $m$ inputs of the channels which are not \emph{confusable} after passing through the channel.
Thus, we can forget about the output set of the channel and only consider its \emph{confusability graph}; that is, a graph on the input set of the channel with two vertices being connected if with non-zero probability the output of the channel is the same under those two inputs. Then encoding $m$ messages into codewords of length one is equivalent to finding an independent set of size $m$ in the graph, and the one-shot zero error capacity of the channel is equal to the logarithm of the independence number of the graph. Moreover, multiple uses of the channel corresponds to tensor product of the graph (defined below) with itself, and then the zero-error capacity is given in terms of the independence number of the tensor powers of the graph.

It is known that computing the independence number of graphs is an $\NP$-complete problem; thus estimating the zero-error capacity is a much harder problem. Even for small graphs such as the cycle of length seven the capacity is unknown. Shannon \cite{shannon} computed the capacity of all graphs up to four vertices, but the capacity of the cycle of length five remained an open problem until Lov\'{a}sz \cite{lov} introduced an upper bound on the zero-error capacity. Via semidefinite programming (SDP) relaxation he found an upper bound on the independence number of graphs. Then using the primal-dual framework of SDP's he showed that the upper bound is multiplicative, and concluded that this is indeed an upper bound on the zero-error capacity.

%*************************************************

\begin{figure}

\begin{center}

\begin{tikzpicture}[scale=.5]

\node at (1, 0) {$x_1$};
%\draw[fill] ( 1.5, 0) circle (.1) ;

\node at (1, -1.5) {$x_2$};
%\draw[fill] ( 1.5, -1.5) circle (.1) ;

\node at (1, -3) {$x_3$};
%\draw[fill] ( 1.5, -3) circle (.1) ;

\node at (1, -4.5) {$x_4$};
%\draw[fill] ( 1.5, -4.5) circle (.1) ;

\node at (1, -6) {$x_5$};
%\draw[fill] ( 1.5, -6) circle (.1) ;

%****

\node at (5.5, -1.5) {$y_1$};
%\draw[fill] ( 4, -1.5) circle (.1) ;

\node at (5.5, -3.75) {$y_2$};
%\draw[fill] ( 4, -3.75) circle (.1) ;

\node at (5.5, -5.25) {$y_3$};
%\draw[fill] ( 4, -5.25) circle (.1) ;

%****

\draw[-] (1.5, 0) -- (5, -1.5);
\node at (3.5, -.5) {\scriptsize 1};

\draw[-] (1.5, -1.5) -- (5, -1.5);
\node at (2.4, -1.15) {\scriptsize 1};

\draw[-] (1.5, -3) -- (5, -1.5);
\node at (2.4, -2.1) {\scriptsize 0.4};

\draw[-] (1.5, -3) -- (5, -3.75);
\node at (3.5, -3) {\scriptsize 0.6};

\draw[-] (1.5, -4.5) -- (5, -3.75);
\node at (2.6, -3.9) {\scriptsize 0.7};

\draw[-] (1.5, -4.5) -- (5, -5.25);
\node at (3.6, -4.6) {\scriptsize 0.3};

\draw[-] (1.5, -6) -- (5, -5.25);
\node at (2.6, -5.4) {\scriptsize 1};

 %*********

\node at (9, -2) {$x_1$};
\draw[fill] ( 9.5, -2) circle (.1) ;

\node at (11.75, -.25) {$x_2$};
\draw[fill] ( 11.75, -.75) circle (.1) ;

\node at (14.6, -2) {$x_3$};
\draw[fill] ( 14, -2) circle (.1) ;

\node at (13.17, -5.2) {$x_4$};
\draw[fill] ( 13.17, -4.8) circle (.1) ;

\node at (10.42, -5.2) {$x_5$};
\draw[fill] ( 10.42, -4.8) circle (.1) ;

%********

\draw[-] (9.5, -2) -- (11.75, -.75);
\draw[-] (9.5, -2) -- (14, -2);
\draw[-] (11.75, -.75) -- (14, -2);
\draw[-] (14, -2) -- (13.17, -4.8);
\draw[-] (10.42, -4.8) -- (13.17, -4.8);

%*********

\end{tikzpicture} \caption{The left graph shows a channel with input set $\{x_1, \dots, x_5\}$ and output set $\{ y_1, y_2, y_3\}$. If, for example, the input of the channel is $x_3$ the output is $y_1$ or $y_2$ with probabilities $0.4$ and $0.6$ respectively. If the receiver sees $y_2$ as the output, the input is either $x_3$ or $x_4$, so $(x_3, x_4)$ is a confusable pair. The right graph depicts the confusability graph of the channel. The independence number of this graph is $2$, so $\log 2=1$ bit of information, by using codewords $\{x_3, x_5\}$, can be sent through this channel with no error. Moreover, it is not hard to convince oneself that the independence of the $k$-th tensor power of this graph (the graph corresponding to $k$ use of the channel) is equal to $2^k$, and then the zero-error capacity of this channel is $1$.    } 
  \label{fig:channel}

\end{center}

\end{figure}
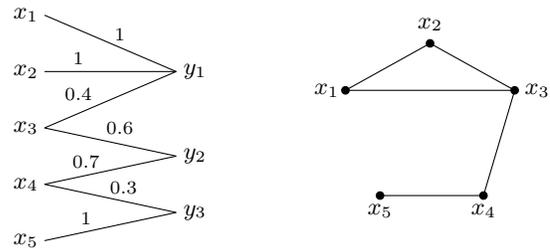
%**************************************************

The entanglement-assisted zero-error capacity of classical channels was recently studied in \cite{ent-zero}; that is, what happens if the sender and receiver can share an entangled state. The authors provided a graph for which the entanglement-assisted one-shot capacity is greater than the independence number of the graph. This example is evidence that unlike the usual capacity we may increase the zero-error capacity by allowing entanglement as a resource. This example, however, deals only with the one-shot capacity, and computing the entanglement-assisted zero-error capacity seems even harder than the graph capacity.

In this paper, we show that the Lov\'{a}sz upper bound on the zero-error capacity is also an upper bound on the entanglement-assisted zero-error capacity. \\

\noindent{\bf Related works.} After finishing this work we found that in an independent work Runyao Duan, Simone Severini, and Andreas Winter \cite{winter} have defined a generalization of the Lov\'{a}sz bound for quantum channels, and have shown that it is an upper bound on the entanglement-assisted zero-error capacity of quantum channels. For classical channels this bound coincides with the Lov\'{a}sz bound and then their work gives our result by a different approach. Besides being shorter, an advantage of our proof is that it gives the intuition of how to
construct an entanglement-assisted communication protocol from the optimal vectors in the Lov\'{a}sz bound.

\section{Entanglement-assisted zero-error capacity}\label{sec:def}

A discrete memoryless (classical) channel consists of a finite input set $X$ and an output set $Y$ and the set of probabilities $p(x\vert y)$ for every $x\in X, y\in Y$, meaning that under the input $x$ the output of the channel is $y$ with probability $p(x\vert y)$. The confusability graph of this channel is a graph $G$ on the vertex set $X$ in which two vertices $x, x'\in X$ are adjacent if there exists $y\in Y$ such that $p(x\vert y)$ and $p(x'\vert y)$ are both non-zero. Inputs $x_1, \dots, x_m$ can encode $m$ messages with zero-error if they are not confusable after passing through the channel, and then the one-shot zero-error capacity of the channel is equal to $\log \alpha(G)$ where $\alpha(G)$ denotes the independence number of $G$, i.e. the maximum number of vertices no two of which are adjacent. The graph corresponding to $k$ uses of the channel is given by the $k$-fold tensor product of $G$ with itself; that is, $G^{\otimes k}$ is a graph on the vertex set $X^{k}$ and there is an edge between two vertices $(x_1,\dots, x_k)$ and $(x'_1,\dots ,x'_k)$ if for each $1\leq i\leq k$, $x_i$ and $x'_i$ are either equal or adjacent in $G$. As a result, the zero-error capacity is given by $\log (\lim \alpha(G^{\otimes k})^{1/k})$ as $k\rightarrow\infty$.

In the presence of entanglement, the sender (Alice) and receiver (Bob) can share an entangled state. To send a message, Alice may apply some local measurement (which depends on the message) in order to decide about the input of the channel. Then Bob after receiving the output applies a measurement (which depends on the output) to find Alice's message. Being a zero-error protocol, Bob's residual states for  different messages, but a fixed output $y\in Y$, should be distinguishable. However, note that Bob's residual state depends on Alice's measurement and not directly on $y$ which is not known to Alice. So elements of $Y$ do not appear in the orthogonality relations that the residual states should satisfy. As a result, to compute the entanglement-assisted zero-error capacity we can again ignore the output set $Y$ and only consider the confusability graph $G$ (see \cite{ent-zero} for more details). 

This observation allows us to assume that the channel has the following special form: the output set of the channel is the edge set of the graph, and under the input $x$ we receive one of the edges connected to $x$ as the output (say) all with equal probability. (For an isolated vertex $x$ we also put $x$ in the output set.) Although this assumption is not crucial and the proof can be given for an arbitrary channel, it simplifies the presentation.

Assume that the one-shot entanglement-assisted zero-error capacity of the channel is $ \log m$. Then the encoding and decoding processes are as follows. Alice and Bob share a bipartite state $\vert \psi\rangle$; to transmit the message $1\leq i\leq m$, Alice measures her part of $\vert \psi\rangle$ using the POVM $\{P_x^i:\, x\in X\}$, and sends the outcome of the measurement $x\in X$ through the channel. The output of the channel is an edge $xx'$ connected to $x$. Then Bob measures the other part of $\vert \psi\rangle$ using the POVM $\{Q_j^{xx'}:\, 1\leq j\leq m\}$ and considers the outcome of this measurement as Alice's message. Since this protocol has no error, the outcome of
Bob's measurement must coincide with Alice's message $i$. That is, for every edge $xx'$
\begin{equation}\label{eq:ortho}
\langle \psi \vert P_x^i\otimes Q^{xx'}_j\vert \psi\rangle = \delta_{ij},
\end{equation}
where $\delta_{ij}$ denotes the Kronecker delta function. In case $x$ is an isolated vertex, Bob's measurement is of the form $\{Q_j^x:\, 1\leq j\leq m\}$ and we should have $\langle \psi \vert P_x^i\otimes Q^{x}_j\vert \psi\rangle = \delta_{ij}$.

As a result, the one-shot entanglement-assisted zero-error capacity of the channel is equal to $\log \alpha^{\ast}(G)$ where $\alpha^{\ast}(G)$ is the maximum number $m$ for which there exist POVMs $\{P_x^i:\, x\in X\}$ and $\{Q_j^{xx'}:\, 1\leq j\leq m\}$ satisfying \eqref{eq:ortho}. Also, the entanglement-assisted zero-error capacity is equal to $\log (\lim \alpha^{\ast}(G^{\otimes k})^{1/k})$ as $k\rightarrow\infty$.

\section{The Lov\'{a}sz theta function}

There are several characterizations of the Lov\'{a}sz theta function \cite{lov}. The following definition is appropriate for us.
\begin{eqnarray*}
\vartheta(G)  & =   \max \tr(BJ) & \\
              &   B \geq 0       &\\
              & \tr B=1         & \\
              &  B_{xx'}=0       & \text{for every edge $xx'$}
\end{eqnarray*}
where $B$ is an $\vert X\vert\times \vert X\vert$ matrix and $J$ is the matrix with all entries equal to one. Considering an independent set of size $\alpha(G)$, and letting $B_{xx'}$ be equal to $1/(\alpha(G))$ for every $x,x'$ in the independent set and $0$ otherwise, we find that $\alpha(G)\leq \vartheta(G)$. Lov\'{a}sz showed that $\vartheta(G)$ is multiplicative ($\vartheta(G\otimes H) = \vartheta(G)\vartheta(H)$) and concluded $\alpha(G^{\otimes k})^{1/k} \leq \vartheta(G)$ and that $\log \vartheta(G)$ is an upper bound on the zero-error capacity of $G$.

Lov\'{a}sz using this upper bound computed the capacity of $C_5$, the cycle of length five. If we let the vertices of $C_5$ be $x_1, \dots, x_5$, $\{x_1x_1, x_2x_3, x_3x_5, x_4x_2, x_5x_4 \}$ is an independent set in $C_5\otimes C_5$. Then the capacity of $C_5$ is at least $\log \sqrt{5} $. On the other hand,  $\vartheta(C_5)= \sqrt{5}$ \cite{lov}. Therefore, the capacity of $C_5$ is equal to $\log \sqrt{5}$.

\section{Main result}

\begin{thm} \label{thm:main} $\vartheta(G) \geq \alpha^{\ast}(G)$, and since $\vartheta(G)$ is multiplicative, $\log \vartheta(G)$ is an upper bound on the entanglement-assisted zero-error capacity of a classical channel with the confusability graph $G$.
\end{thm}

Let us first explain the idea behind the proof of this theorem  and then go through the details. Assume that $\alpha^{\ast}(G) = m$ and for every $1 \leq i\leq m$ and $x\in X$ there exists a vector $w_x^i$ in some inner product vector space satisfying the following conditions:
\begin{itemize}
\item[(1)] for every $i$, $\sum_x w_x^i = w$ where $w$ is a vector of length one,
\item[(2)] $\langle w_x^i , w_{x'}^i\rangle =0$ for every $x\neq x'$,
\item[(3)] $\langle w_x^i, w_{x'}^j\rangle =0 $ for every edge $xx'$ of $G$,
\item[(4)] and $\langle w_x^i, w_x^j \rangle =0$ for every $i\neq j$.
\end{itemize}
Then we can define the $\vert X\vert \times \vert X\vert $ matrix $B$ by $B_{xx'} = \langle w_x, w_{x'}\rangle$ where $w_x = \sum_i w_x^i$. In this case, $B$ is a positive semidefinite matrix and one can show that $B_{xx'}=0$ for every edge $xx'$, $\tr B = m$ and $\tr (BJ)= m^2$ (later we will prove them all). Then by the definition of $\vartheta(G)$ we have $\vartheta(G)\geq \tr(BJ)/ \tr B = m = \alpha^{\ast} (G)$. So our main problem is to define the vectors $w_x^i$ satisfying the above properties. In the case where the shared state between Alice and Bob is the maximally entangled state, these vectors are basically the residual states of Alice after her measurement, but the general case needs more work because we should define some twisted inner product.

Assume that $m=\alpha^{\ast}(G)$ and there exist a bipartite state $\vert \psi\rangle$ and POVM measurements $\{P_x^i:\, x\in X\}$ for $1\leq i\leq m$, and $\{Q_j^{xx'}:\, 1\leq j\leq m\}$ for every edge $xx'$, satisfying \eqref{eq:ortho}. Due to the normalization of the POVMs we have
$$\sum_{x\in X} P_x^i = I$$
and
$$\sum_{j=1}^{m} Q_{j}^{xx'} =I.$$
Without loss of generality, we assume that the POVMs $\{P_x^i:\, x\in X\}$ are projective measurements ($P_x^iP_{x'}^i = \delta_{x,x'} P_x^i$) because every POVM measurement can be written as a projective measurement on an extended space. Also, we may assume that the local spaces of Alice and Bob are isomorphic and of dimension $d$. Let $\vert \Phi\rangle$ be the maximally entangled state on these two spaces. Then there exists a matrix $S$ such that
$$\vert \psi\rangle = I\otimes S\vert \Phi\rangle,$$
and due to the normalization of $\vert \psi\rangle$ we have $\tr\, S^{\dagger}S=d$.

For every $1\leq j\leq m$ and edge $xx'$ define $R_j^{xx'} = (S^{\dagger} Q_j^{xx'} S)^{\T}$, where $\T$ denotes the transpose operation with respect to the orthonormal basis in which the maximally entangled state $\vert \Phi\rangle$ is defined. Then for every edge $xx'$ we have
$$R_1^{xx'}+\cdots +R_m^{xx'}=M,$$
where $M=(S^{\dagger}S)^{\T}$. Observe that
\begin{eqnarray*}
\delta_{ij} & = & \langle \psi\vert P_x^{i}\otimes Q_j^{xx'}\vert \psi\rangle \nonumber\\
            & = &\langle \Phi \vert P_x^{i}\otimes S^{\dagger}Q_j^{xx'} S\vert \Phi\rangle \nonumber\\
            & = & \frac{1}{d} \tr( P_{x}^{i} R_{j}^{xx'}).
\end{eqnarray*}
Since both $P_x^i$ and $R_j^{xx'}$ are positive semidefinite we conclude that
\begin{equation}\label{eq:main}
P_x^iR_j^{xx'}=0,
\end{equation}
for every edge $xx'$ and every $i\neq j$.

Consider the vector space of complex $d\times d$ matrices equipped with the bilinear form
$$\langle A, B \rangle = \frac{1}{d}\tr (A^{\dagger}BM).$$
Note that, for every non-zero matrix $A$, $\langle A, A\rangle$ is non-negative because $M$ is positive semidefinite. So $\langle \cdot, \cdot\rangle$ is a non-negative form \footnote{It may not be positive because $M$ can have zero eigenvalues.}. There are some orthogonality relations among matrices $P_x^i$ with respect to this bilinear form.

\begin{itemize}
\item[\rm{(a)}] $\langle P_x^i, P_{x'}^i\rangle = 0$ for every $x\neq x'$.

 This is simply because $P_x^iP_{x'}^i=0$.

\item[\rm{(b)}] $\langle P_x^i, P_{x'}^j\rangle = 0$ for every edge $xx'$.

 From \rm{(a)} we may assume that $i\neq j$. Then we have
\begin{equation*}
\langle P_x^i, P_{x'}^j\rangle   = \frac{1}{d} \tr( P_x^i P_{x'}^j M)
 = \frac{1}{d} \sum_{k=1}^m \tr(P_x^i P_{x'}^j R_k^{xx'}) = 0,
\end{equation*}
since for every $k$, by \eqref{eq:main}, either $P_x^i R_k^{xx'}=0$ or $P_{x'}^j R_k^{xx'}=0$.

\item[\rm{(c)}] $\langle P_x^i, P_{x}^j\rangle = 0$ for every $i\neq j$.

Let $x'$ be an adjacent vertex of $x$. As before for every $k$ either $P_x^i R_k^{xx'}=0$ or $P_{x}^j R_k^{xx'}=0$ and orthogonality follows. Similar equations can be written if $x$ is an isolated vertex.
\end{itemize}

Now for every vertex $x$ of $G$ let $V_x= \sum_{i=1}^m P_x^i$ and define the matrix $B$ by $B_{xx'}= \langle V_x, V_{x'}\rangle$. Since $\langle \cdot, \cdot \rangle$ is a non-negative form, $B$ is positive semidefinite. Moreover, by (b) for every edge $xx'$ we have
\begin{align*}
B_{xx'}=\langle V_x , V_{x'}\rangle = \sum_{i,j=1}^m \langle P_x^i, P_{x'}^{j}\rangle =0 .
\end{align*}
As a result, $\tr(BJ)/\tr(B) \leq \vartheta(G)$. So the only remaining part is to show $\tr(BJ)/\tr(B) = m$.
Using (c) we have
\begin{align*}
\tr B & = \sum_{x} \langle V_x, V_x\rangle = \sum_{i,j=1}^m \sum_x\, \langle P_x^i, P_x^{j}\rangle = \sum_{i=1}^m \sum_x \,\langle P_x^i, P_x^i\rangle\\
& = \frac{1}{d} \sum_{i=1}^m \sum_x \, \tr(P_x^i M) = \frac{1}{d} \sum_{i=1}^m \tr(M) =m,
\end{align*}
where in the second line we use $(P_x^i)^2= P_x^i$.
Also by $\sum_x P_x^i=I$ we obtain
\begin{align*}
\tr(BJ) & = \sum_{x,x'} \, \langle V_x, V_{x'}\rangle = \sum_{i,j=1}^m \sum_{x,x'} \,\langle P_x^i, P_{x'}^j\rangle \\\
& = \sum_{i,j}^m \,\langle I, I\rangle = m^2.
\end{align*}
We are done.

\section{Discussion}

The proof of Theorem \ref{thm:main} indeed gives the following stronger statement. Let $\beta(G)$ be the maximum number $m$ such that there exist vectors $w_x^i$, $1\leq i\leq m$, in an inner product vector space satisfying conditions (1)-(4) of the previous section. Then $\alpha^{\ast}(G)\leq \beta(G) \leq \vartheta(G)$. It is a very interesting question whether $\alpha^{\ast}(G)$ and $\beta(G)$ are always equal or not. 
The vectors constructed in the proof of Theorem \ref{thm:main} come from positive semidefinite matrices (POVM elements), but there is no such a constraint in the definition of $\beta(G)$, and this point seems to be the main difference between $\alpha^{\ast}(G)$ and $\beta(G)$. Another interesting problem is the relation between $\beta(G)$ and $\vartheta(G)$; for example, do we have  $\beta(G) = \lfloor \vartheta(G) \rfloor$? To prove this equality, given a matrix $B$ in the definition of $\vartheta(G)$, if $B_{xx'} = \langle w_x, w_{x'}\rangle$, we should be able to decompose each $w_x$ into $\lfloor \vartheta(G) \rfloor$ vectors $w_x = \sum_i w_x^i$ satisfying (1)-(4).  In general, computing $\beta(G)$ seems easier than $\alpha^{\ast}(G)$ and we hope this quantity furthers research in this direction.

Examples of \cite{ent-zero} which show a gap between $\alpha(G)$ and $\alpha^{\ast}(G)$ are defined based on the Kochen-Specker sets. In these examples one can easily check that $\alpha^{\ast}(G)$ and $\vartheta(G)$ coincide, so according to Theorem \ref{thm:main}  the entanglement-assisted zero-error capacity of these channels is equal to $\log \vartheta(G)$. Thus computing the usual zero-error capacity of these graphs and comparing it with the Lov\'{a}sz theta function would clarify the role of entanglement in zero-error communication. Also, finding instances of graphs $G$ with $\alpha(G) < \alpha(G)^{\ast} < \vartheta(G)$ is of interest.

Finally, as noted in \cite{ent-zero} entanglement-assisted zero-error communication protocols are related to pseudo-telepathy games. So techniques of this paper might be useful for studying such games.

\vspace{.5cm}

\noindent{\bf Acknowledgements.} The author is grateful to the unknown referee whose comments helped to improve the presentation of the paper. This work has been supported in part by NSF under Grant No. PHY-0803371
and by NSA/ARO under Grant No. W911NF-09-1-0442.

\end{document}